# Data Infrastructure and Approaches for Ontology-Based Drug Repurposing

Stephen Boyer, Thomas Griffin, Sarath Swaminathan, Kenneth L. Clarkson, Dmitry Zubarev

*IBM Almaden Research Center, 650 Harry Road, San Jose, California 95136*

**Abstract**

We report development of a data infrastructure for drug repurposing that takes advantage of two currently available chemical ontologies. The data infrastructure includes a database of compound-target associations augmented with molecular ontological labels. It also contains two computational tools for prediction of new associations. We describe two drug-repurposing systems: one, *Nascent Ontological Information Retrieval for Drug Repurposing (NOIR-DR)*, based on an information retrieval strategy, and another, based on non-negative matrix factorization together with compound similarity, that was inspired by recommender systems. We report the performance of both tools on a drug-repurposing task.

**Introduction**

Drug repurposing is an efficient strategy for drug discovery, where new targets or activities are found for known drugs [1-5]. Drug repurposing requires the efficient representation of existing information about the activity of chemical compounds as drugs, and the development of algorithms that leverage such information and propose new indications. In this contribution, we develop a computational infrastructure for drug repurposing. It includes a database of compounds decorated with structural ontological labels and methods for prediction of new drug-like activity. The latter methods comprise *Nascent Ontological Information Retrieval for Drug Repurposing (NOIR-DR),* and a recommender engine based on non-negative matrix factorization (NMF). In recent years, chemometric representation of molecular structure has seen expansion via construction of chemical ontologies [6-8]. General ontologies represent human knowledge in a computationally accessible form. Chemical ontologies are motivated by diverse classification goals, such as structural relations between molecules, bioactivity of compounds, or even classification of cheminformatic descriptors and algorithms.

The primary uses of the existing chemical ontologies are classification of database content, fast search and data retrieval, entity resolution, and automated annotation of novel chemicals [7, 8]. We recognize that chemical ontological labels fall into the category of structural keys, that is, data structures such as bitmaps, in which each entry represents the presence or absence of a specific structural feature [9-15]. Work in chemical informatics and machine learning suggests that indeed structural keys or molecular fingerprints can be used for predictive analytics of molecular properties/activities [16-21]. In this work, we want to benchmark performance of chemical ontologies in such predictive tasks.

We have developed a data infrastructure and methods that facilitate drug repurposing via: (a) construction of a database of compound-target associations derived from ChEMBL data [22, 23] augmented with ontological labels derived from ClassyFire [7] and OntoChem [8] chemical ontologies, (b) a database search strategy NOIR-DR, and (c) a recommender system for drug repurposing based on NMF. This work consists of forward form of drug repurposing, where predictions are made about new targets for the specified compound, and its inverse form, where predictions are made about new active compounds for the specified target. Molecular ontological labels are human-interpretable and suitable for building search queries by chemists without expertise in statistical data analysis. NOIR-DR search takes advantage of this distinctive feature of chemical ontologies. Given a target, NOIR-DR strategy is used to retrieve new compounds that are expected to exhibit activity toward the target in question, thus solving the inverse drug-repurposing problem. The recommender system is capable of solving both versions of drug repurposing; here we give details about its application to the forward problem.

**Data and Methods**

*Ontological labels*

Two programs were used to generate the chemical ontology as the source of features for subsequent information retrieval and collaborative filtering steps. One program, which we will call OC, was developed by OntoChem [7], and the second program ClassyFire (or CF, for short) was developed at the University of Alberta [8]. Both programs take in a SMILES string as input and output a list of molecular attributes (chemical labels); see Figure 1.

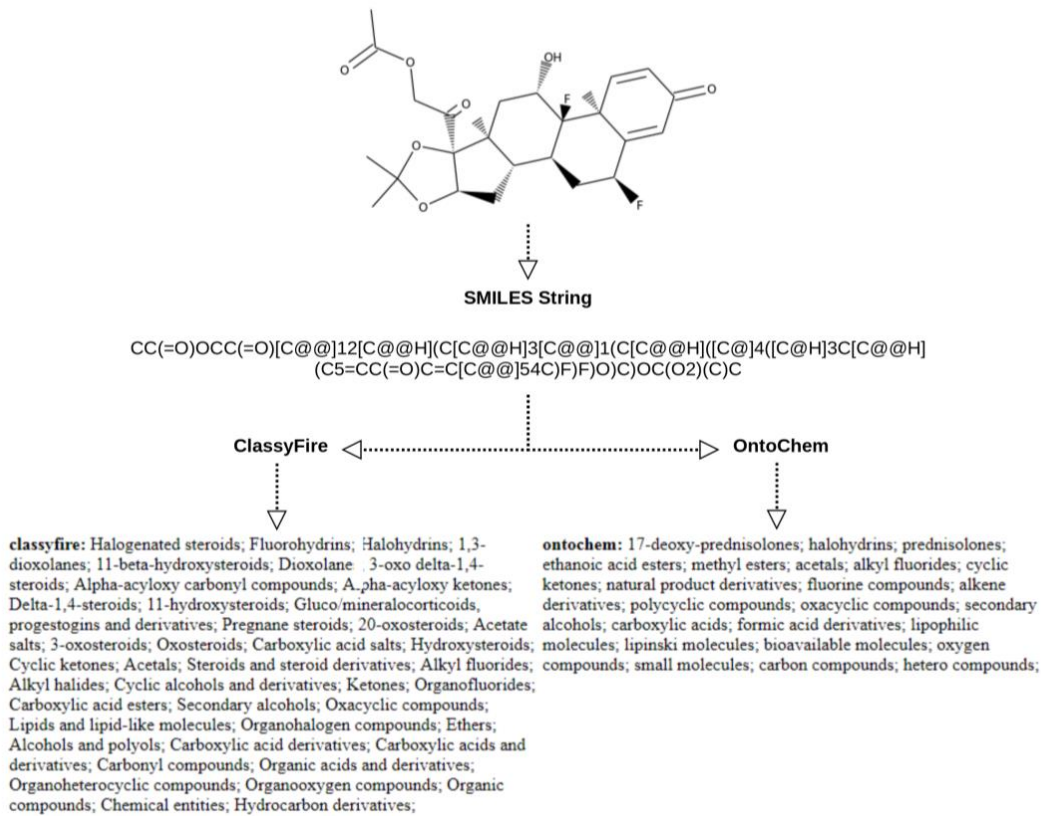

**Figure 1.** Example of ClassyFire and OntoChem ontological labels generated for fluocinonide molecule.

*Database augmentation*

ChEMBL release 22 was used [22]. This database, of roughly 1.4 million compounds together with their known biological activity, has been manually curated for many years. ChEMBL associates small molecules with targets, storing also the corresponding bioactivity value. The ChEMBL database was augmented using the OC and CF programs to generate ontologies for each molecule in the database (Figure 2). We refer to the final form of the database as the `ChEMBL ontology dB`.

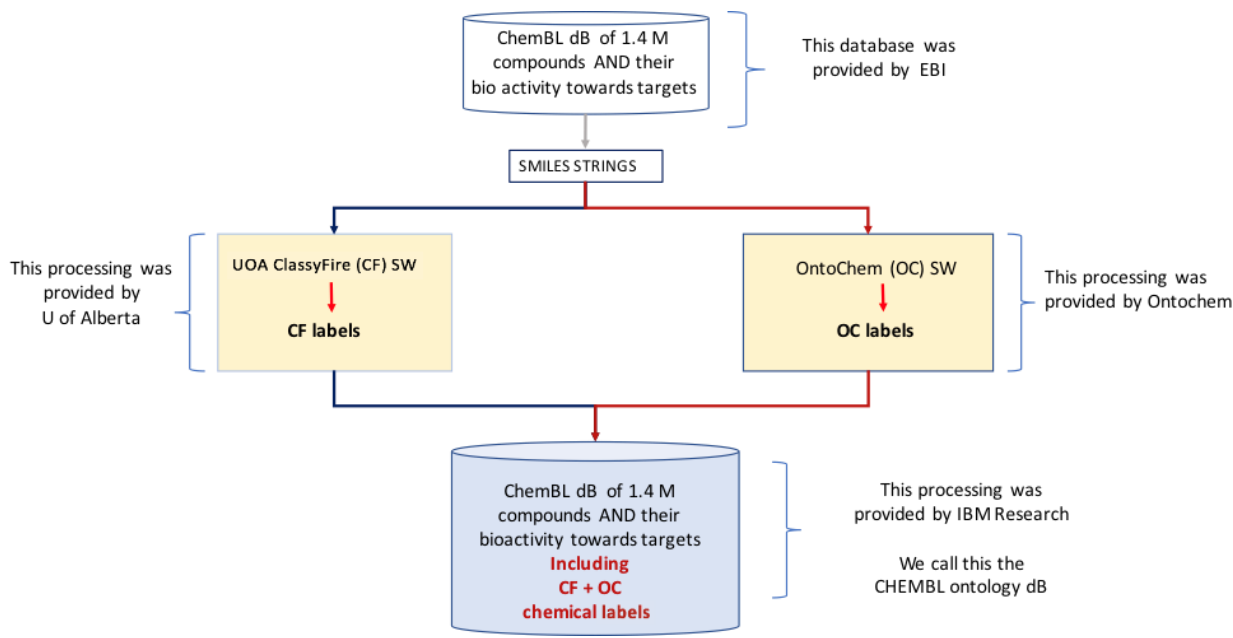

**Figure 2.** Construction of the `ChEMBL ontology dB`.

The two ontology programs (CF and OC) ingest a SMILES string, which is a machine-readable representation of the structure of a molecule, and find various attributes of the molecule. For example, the molecule is identified as a steroid, an olefin, a carboxylic acid. etc. Other attempts have been made to do similar analyses using chemical fingerprints. This has recently been referred to as a "similarity ensemble approach" (SEA). Information regarding SEA analysis can be found in [17].

Once the chemical labels have been generated, the ontological attributes are loaded into what becomes the enhanced version of the ChEMBL database that now includes the ChEMBL molecules, their molecular attributes (chemical ontological labels), the targets associated with those molecules, and assay information for the compound-target relationship.

Having decorated the ChEMBL data with chemical ontological labels, the associations between compounds and biochemical targets are predicted. The following section describes two drug-repurposing strategies, one inspired by information retrieval methods, and another inspired by recommendation systems.

*Nascent Ontological Information Retrieval for Drug Repurposing (NOIR-DR)*

Among existing definitions of structural keys, fingerprints, and other means of expression of molecular structure in a chemometric context, chemical ontologies stand out because they are naturally interpretable by subject matter experts. Ontological labels can be used to query chemical databases via natural language, which is the preferred form of interaction with the data for non-experts in cheminformatics and machine learning. Also, ontological labels can be trivially plugged into the standard workflows of feature engineering and supervised/unsupervised learning in chemometrics.

The first form of drug repurposing we describe is an information retrieval approach that takes advantage of the ontology-augmented database (Fig. 1). Recently, a method of generating scientific hypotheses based on a semantic "bag of words" embedding of the document content was proposed [24]. Along with predictions of activity of several kinases in targeting p53 tumor suppressor protein, the paper reported the experimental validation of the activity of several predicted p53 kinases. In this approach, the following roles are assigned: the set of compounds in the `ChEMBL ontology dB` is a corpus, compounds are documents, and molecular attributes, such as ontological labels, are terms in the documents. Biochemical targets are treated as topics of the documents. The information retrieval task is to extract the documents that are relevant to the specific topic based on keywords associated with the topic. Selection of the keywords for the query can be motivated by external considerations, and using a broad range of statistical analytic tools that elucidate term-topic associations. For illustrative purposes, a set of ontological labels indicative of the desired compound activity is selected on the basis of a simple *ad-hoc* scoring of terms motivated by $\chi^2$ statistics of term counts, as follows.

$$TermScore_i = (O_i - E_i)^2 / E_i \tag{1}$$

$$E_i = C_i \times N_{relevant} / N_{corpus} \tag{1a}$$

Here:
$TermScore_i$ is the score of the search term, i.e., ontological label $i$;
$O_i$ is the observed count of the term in the documents that are known to be relevant to the search; and
$E_i$ is the expected count of the term based on the occurrence of the term in the entire corpus.

$E_i$ is computed according to Eq. 1a, where:
$C_i$ is the count of the term $i$ over the entire corpus;
$N_{relevant}$ is the count of the documents that are known to be relevant to the search; and
$N_{corpus}$ is the number of documents in the corpus.

The documents are scored according to Equation 2.

$$DocScore_i = \frac{1}{L_i} \sum_{j < L_i} TermScore_j^i \tag{2}$$

where $DocScore_i$ is the score of the document, i.e., molecule $i$, $L_i$ is the number of terms found in the document, and $TermScore_j^i$ is the score of term $j$ found in the document $i$.

*Matrix Factorization for Compound-Target Association*

Another possibility is to approach the prediction of compound-target associations as an instance of a recommendation problem, and use the machinery of collaborative filtering. In collaborative filtering, recommendations are predictions of new associations, such as users with preferred movies, shoppers with products, etc. made on the basis of the existing associations. In the simplest form, this can be done without any information about the similarity of the involved objects.

Matrix factorization (MF) is a class of algorithms that enables collaborative filtering. The conceptual premise of MF is the existence of a few latent factors that explain associations. In recommendations of movies, for example, a latent factor might be the genre of a movie and the user preference for a specific genre. In compound-target associations, one interpretation of the latent factors is the mechanism of drug activity, that is, the mode of interaction between compounds and targets.

Various matrix factorization approaches have been explored over the past decade for predicting novel drug-target interactions (DTI). Gonen et al [26] presented one of the first works utilizing matrix factorization for DTI prediction, using a Bayesian formulation combining dimensionality reduction, matrix factorization and binary classification. Cobanoglu et al [27] used Active Learning along with Probabilistic Matrix Factorization to predict novel drug-target interactions. Zheng et al [28] incorporated multiple drug-drug and target-target similarity measures with an Alternating Least Squares algorithm for matrix factorization to predict DTI. Liu et al. [29] formulated a Logistic Matrix Factorization model regularized by neighborhood information of drugs and targets. Ezzat et al. [30] proposed a Graph Regularized Matrix Factorization approach which also makes use of drug-drug and target-target similarity information to improve the accuracy.

In its simplest form, MF does not involve any evaluation of the structural similarity of the objects. The algorithm can be enhanced along these lines with the hope of improving the prediction quality. Incorporation of compound features and similarity metrics into MF algorithms also offers a solution to the cold-start problem encountered when a recommendation has to be offered for a compound that has no association data available in the dataset.

In this study, we consider two forms of Nonnegative Matrix Factorization (NMF): the basic one, without evaluation of pairwise similarity of compounds, and an enhanced one, where similarity evaluation for pairs of compounds contributes to the prediction of the compound-target associations. NMF is a family of MF algorithms which is suitable for sparse matrices with non-negative values. NMF algorithms factorize the data matrix *X* to latent feature matrices *U* and *V*, with the constraint that $U$ and $V$ have no negative values. This non-negativity property of NMF algorithms allows them to be interpretable and suitable for clustering.

NMF learns $U$ and $V$ that together minimize the following objective function:

$$J = \frac{1}{2}\|X - UV^T\|^2, \text{ subject to } U \geq 0, V \geq 0 \qquad (3)$$

A multiplicative update rule that can be used to minimize this objective function was proposed in [31].

When learning a model for predicting drug-target interactions, the basic NMF algorithm takes only the known interactions into account. Following the intuition that structurally similar compounds might be active against the same target, we can incorporate compound similarity information with NMF to enable a better model and improve recommendation accuracy. A second MF approach, based on [25], uses compound similarity information to force the latent feature representation of similar compounds to be closer to each other (compound similarity NMF or CS-NMF).

Suppose we are given a pairwise compound similarity matrix $S$, where $S_{ij}$ is the similarity or the thresholded similarity between compounds $i$ and $j$. CS-NMF learns $U$ and $V$ that minimizes the following objective function:

$$J = \frac{1}{2}\|X - UV^T\|^2 + \frac{\lambda}{2}\sum_{i,j} S_{ij}\|u_i - u_j\|^2, \text{ subject to } U \geq 0, V \geq 0 \qquad (4)$$

Refer to [25] for the detailed solution of this objective function.

The structural similarity between compounds can be calculated using different ways based on the representation of compound and similarity metrics used. Here we use the Morgan fingerprint and the two ontological fingerprints Ontochem and ClassyFire to represent the compounds. Similarity between compounds is measured using Jaccard Similarity.

**Results and Discussion**

*NOIR-DR*

As an example, a search query and information retrieval scheme is constructed for the target *CHEMBL1907611 "Tumour suppressor p53/oncoprotein Mdm2"*, or MDM2 for short. This target is an oncogene important for the treatment of cancer. The `ChEMBL ontology dB` database is queried for all known compounds with biological activity for this target. ChEMBL22, the version of the ChEMBL database used in this study, had a total of 1.4 million molecules of which 585 molecules had known activity for MDM2.

The first task of NOIR-DR is to identify a reference set of ontological labels that can be used for the construction of human-interpretable queries for the retrieval of compounds with possible activity toward MDM2 from our database. The relevant class of 585 active compounds is narrowed down by introducing an activity threshold so that a subset of compounds is selected whose effective concentration of biological activity (EC50) is less than 30 nM. Other structure-activity relationships, such as lethal dose (LD50), could have been selected for toxicological activity. In addition to the activity threshold, construction of the reference set of ontological labels involves selection of two more parameters. One is the level of noise, which is the highest allowed count for a label over the entire corpus. Labels are discarded whose observed corpus count exceeds 200K. Another parameter is the size of the reference set, in this example limited to 20 labels.

The number of labels that appear more than once among 44 high-activity compounds and do not exceed the selected level of noise is 130 for ClassyFire and 122 for OntoChem. For each ontology, the top-20 ranking labels are selected according to their computed score (Eq. 1) and two reference label sets are constructed. The members of these label sets with their respective scores are shown in Table 1.

Among compounds with known activity against MDM2 that do not pass activity threshold, at least one of the reference labels from ClassyFire ontology is present in 393 compounds, and at least one of the reference labels from OntoChem ontology is present in 314 compounds. Overall, 242 of the known active compounds that do not pass the activity threshold carry at least one ClassyFire and at least one OntoChem reference label. A consensus between the reference label sets derived from the two ontologies is, therefore, achieved on 41% of compounds in question.

In the information retrieval step, the entire corpus of `ChEMBL ontology dB` compounds minus 44 high-activity compounds is searched against the query containing the ontological labels from the reference set. The compound score is computed according to Eq. 2. Top 100 scoring compounds per each ontology are selected. At this point, retrieved data include some of the compounds that are known as active against MDM2 per ChEMBL version used in the study, along with compounds that are not known to be active against MDM2 per used version of ChEMBL.

**Table 1.** Ontological labels per ClassyFire and OntoChem ontologies that form the reference label set and are used to construct search queries for retrieval of compounds relevant to activity against MDM2 target.

| Label, ClassyFire | Score | Label, OntoChem | Score |
| --- | --- | --- | --- |
| 1-phenyltetrahydroisoquinolines | 15277 | imidazo[1,2-b]thiazoles | 2946 |
| Imidazothiazoles | 8505 | hydroisoquinoline derivatives | 1073 |
| Isoquinolones and derivatives | 1151 | tetrahydroisoquinolines | 1073 |
| Stilbenes | 1151 | 36-membered macrocycles | 834 |
| Thiazolines | 1062 | isoquinolines | 666 |
| Imidazolines | 559 | lactams | 377 |
| Isothioureas | 559 | isothioureas | 374 |
| Tetrahydroisoquinolines | 553 | cyclohexylamines | 352 |
| Thiazolecarboxylic acids and derivatives | 544 | 5,5-membered heterocycles | 243 |
| Isoquinolines and derivatives | 506 | methylamines | 86 |
| Delta lactams | 373 | 3-aminopyridines | 76 |
| Cyclohexylamines | 354 | cyclic amides | 57 |
| Piperidinones | 301 | lactams | 57 |
| Enamines | 266 | alpha-amino acid derivatives | 51 |
| Chlorobenzenes | 247 | amino acid derivatives | 51 |
| Pyrrolidinecarboxamides | 183 | 31-40-membered macrocycles | 51 |
| N-methylpiperazines | 161 | alkylamines | 48 |
| Phenylpiperidines | 148 | threonine derivatives | 42 |
| Pyrrolidine carboxylic acids and derivatives | 145 | tertiary mixed amines | 42 |
| Azolines | 140 | serine derivatives | 31 |

These two retrieved sets have 57 molecules in common, constituting a consensus set. Out of 57 consensus compounds, 9 are known to be active toward MDM2 but did not pass selection thresholds and hence were not included into the training set. The remaining 48 consensus compounds proceed to validation of their activity from external sources. A viable strategy here is a retrospective study [24][1]. Out of 48 consensus compounds in question, 15 (Fig. 3) have confirmed activity toward human MDM2 according to ChEMBL release 23.

---

[1] Our "ChEMBL ontology dB" incorporates compound-target associations according to release 22 of ChEMBL database [22]. Currently, release 23 of ChEMBL is available [23]; we expect it to contain information about compound-target associations that either were discovered between ChEMBL releases or took some time to propagate into the database from the literature.

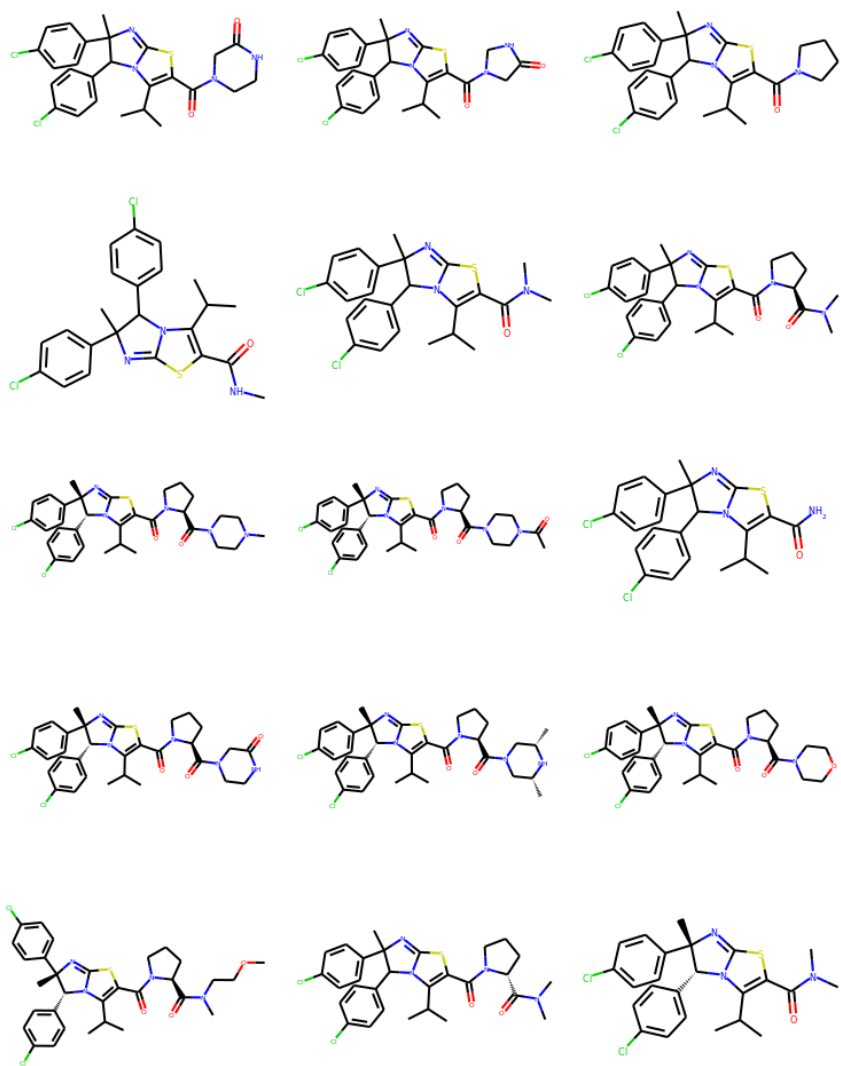

**Figure 3.** Consensus compounds whose predicted activity toward MDM2 is confirmed in the retrospective study. All these compounds can be tracked down to US patent US 2009/0312310 A1.

**NMF and CS-NMF**

In the experiments described in this section a subset of ChEMBL22 is used for training and cross-validation of NMF and CS-NMF. Compound-target interactions were considered only for the targets associated with *Homo Sapiens*. Only compounds and targets with at least one activity reported in the database were selected. The dataset obtained in this manner included 669037 compounds associated with 3381 targets. Though a primary goal is a reliable prediction of the interactions with IC50 less than 10,000 nM, we retained interactions with higher IC50 values in the dataset. The IC50 values are represented in the compound-target adjacency matrix as follows:

- Interactions with IC50 greater than 10,000 nM are represented in data matrix by 1
- IC50 values 0 to 10,000 nM are linearly transformed to [10, 5], by (20,000 – IC50) / 2000. This linear transformation assures that smaller IC50 values are represented by higher values in the matrix.

We evaluated the algorithm in 5-fold cross validation with the following performance metrics:
- RMSE: Root Mean Square Error on test data
- Recall-at-$k$: Recall-at-$k$ is the recall on top k predictions and measured as:
    - Take 10,000 compounds at random from test data which have at least 3 targets in the training data and 3 targets in the test data
    - For each compound, predict score for all targets
    - Rank targets based on predicted score and take top k targets
    - Measure recall on top k targets
    - Measure mean and standard deviation of recall on the 10,000 test compounds

Results for these performance metrics are shown in Table 2 and Figure 3. Performance of CS-NMF improves in comparison to NMF in terms of both RMSE and Recall (Table 2). The Rank – Recall plot shows the fraction of known test targets. In Figure 3, around 85% of the test targets are present in the top 50 target predictions and more than 90% are present in the top 100 target predictions.

This result indicates that the information about structural similarity of compounds helps to provide better recommendations regarding drug-target interaction. The modest degree of improvement suggests a need for deeper investigation of CS-NMF algorithms in the selection of the objective function and regularization technique with the goal of increasing the performance gain.

**Table 2**. Performance metrics for NMF and CS-NMF.

|  | NMF | CS-NMF (Morgan Fingerprint) | CS-NMF (OntoChem) | CS-NMF (ClassyFire) |
|---|---|---|---|---|
| RMSE | 3.95 | 3.93 | 3.93 | 3.93 |
| Recall at 30 | 0.75 (0.31) | 0.76 (0.31) | 0.76 (0.31) | 0.76 (0.31) |
| Recall at 50 | 0.84 (0.27) | 0.85 (0.26) | 0.85 (0.26) | 0.85 (0.26) |

The performance metrics also indicate that CS-NMF performs equally well irrespective of the type of fingerprints used. The average Root Mean Squared Error (RMSE) and Recall on test data were the same when the compound similarity information from Morgan fingerprint, OntoChem and ClassyFire were used. This result indicates that the current version of the NMF algorithm is relatively robust with respect to the introduction of feature-based similarity into the optimization process. Further development is warranted of specialized NMF algorithms that take full advantage of the availability of structural features.

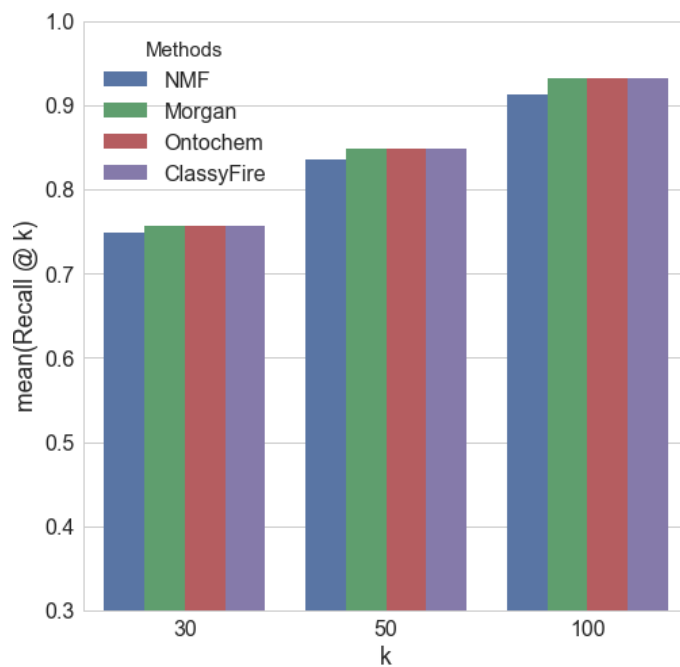

**Figure 3**. Rank – Recall plot for the top 30, 50 and 100 target predictions

Finally, a comparison is provided of the Recommender system for drug repurposing, using NMF and CS-NMF with three types of fingerprints and available predicted associations from ChEMBL. Predictions are made for a sample of 10 existing drugs available on market. The reported numbers are counts of true positive predictions in top-30 predicted targets.

**Table 3.** The number of true positive predictions from our repurposing methodologies and other available sources of predictions (columns). Top 30 predictions are considered for 10 sample drugs (rows).

|  | **NMF** | **CS-NMF Morgan** | **CS-NMF OntoChem** | **CS-NMF ClassyFire** | **ChEMBL** |
|---|---|---|---|---|---|
| ALPELISIB BYL719 PI3K inhibitor | 7 | 7 | 7 | 7 | 23 |
| AMITRIPTYLINE [CHEMBL629] | 27 | 27 | 27 | 27 | 173 |
| DAPAGLIFLOZIN [CHEMBL429910] | 2 | 0 | 0 | 0 | 8 |
| BEPRIDIL [CHEMBL1008] | 13 | 18 | 18 | 18 | 51 |
| SITAGLIPTIN [CHEMBL1422] | 6 | 4 | 4 | 4 | 15 |
| SUMATRIPTAN [CHEMBL128] | 21 | 23 | 23 | 23 | 129 |
| METHFORMIN [CHEMBL1431] | 11 | 12 | 12 | 12 | 129 |
| GLIPIZIDE [CHEMBL1073] | 13 | 16 | 16 | 16 | 124 |
| VALSARTAN [CHEMBL1069] | 18 | 12 | 12 | 12 | 113 |
| ROPINIROLE [CHEMBL589] | 10 | 15 | 15 | 15 | 30 |
| **Total** | 128 | 134 | 134 | 134 | 795 |

**Conclusion**

An infrastructure for data-driven drug repurposing has been developed that includes a compound-target association database augmented with ontological labels for compounds. It facilitates two approaches to the prediction of new associations, an information retrieval NOIR-DR and an NMF-based recommender system.

In information retrieval, human interpretability of ontological labels allows a user without expertise in cheminformatics to organize a retrieval process and to arrive at valuable predictions. The example described here captures a real use-case of the query construction and a database search where the chemist user is able to identify 48 new compounds as potentially active against MDM2, to and confirm activity of 15 of them via retrospective methodology. This example is extremely encouraging, because it bridges the gap between the data-analytic/machine learning community and chemists as subject matter experts. The factor of human interpretability of chemical ontological labels is indispensable in enabling human-driven information retrieval. A supervised learning version of NOIR-DR, such as a binary classification of compounds in an ontological vector space model can be implemented following the methodology of the Knowledge Integration Toolkit [24], for example. It is worth noting that two available ontologies, ClassyFire and OntoChem, showed a modest degree of consensus in labeling reference data and retrieving new data. This is hardly surprising, given that neither ontology is specifically designed to be reflective of the chemical knowledge in the specific domain of anti-cancer drug activity. Development of domain-specific semantic assets, such as ontologies, would be desirable to improve information retrieval performance.

In addressing the applicability of chemical ontologies for drug repurposing via machine learning, (such as the NMF-based recommender system), two forms of NMF are benchmarked, with and without compound similarity evaluation. Introduction of structural features both in the form of Morgan structural fingerprints and ontological labels led to the improvement of NMF performance metrics. Differences in performance were not observed between the effect of structural fingerprints and ontological labels nor between performance of either of the two types of ontological labels. The full potential of CS-NMF in combination with ontological labels will be realized on development of specialized objective functions and regularization techniques along with problem-specific ontologies.